\documentstyle[11pt,IAU207_pasp,twoside,psfig]{article}
\def\edcomment#1{\iffalse\marginpar{\raggedright\sl#1\/}\else\relax\fi}
\reversemarginpar

\newcommand{\Msun}{\mbox{$M_\odot$}}

\newcommand{\comment}[1]{}

\begin{document}
\title{Broad-band photometric evolution of star clusters}
\author{L\'eo Girardi}
\affil{Dipartimento di Astronomia, Universit\`a di Padova\\
           Vicolo dell'Osservatorio 5, I-35122 Padova, Italy}
\affil{Max-Planck-Institut f\"ur Astrophysik,
	Karl-Schwarzschild-Str.~1, D-85741 Garching bei M\"unchen, 
	Germany}

\begin{abstract}
I briefly introduce a database of models that describe the evolution
of star clusters in several broad-band photometric systems. Models are
based on the latest Padova stellar evolutionary tracks -- now
including the $\alpha$-enhanced case and improved AGB models -- and a
revised library of synthetic spectra from model atmospheres.  As of
today, we have revised isochrones in Johnson-Cousins-Glass, HST/WFPC2,
HST/NICMOS, Thuan-Gunn, and Washington systems. Several other filter
sets are included in a preliminary way, like those used by the EIS and
SDSS projects. The database contains also integrated magnitudes of
single-burst stellar populations and Monte-Carlo simulations that show
the stochastic dispersion of the colours as a function of cluster
mass, age, and metallicity. The models are useful for several kinds of
studies, including estimates of masses and ages of extragalactic star
clusters observed by means of broad-band photometry.
\end{abstract}
\keywords{stars -- clusters -- broad-band photometry}

\section{Introduction}
\label{sec_intro}

Broad-band magnitudes and colours are the primary source of
information about the masses and ages of star clusters outside the
Local Group.  Up to recently, this was true even for the relatively
nearby Magellanic Cloud (MC) clusters.

The simplest method of estimating ages and masses comes from the
direct comparison of the observed colours and magnitudes with those
predicted by population synthesis models: Two-colour diagrams may be
used to assign an approximate age to a cluster; then, the integrated
magnitude can be converted into a mass by means of the mass-to-light
ratio predicted for that age. Of course, using models with complete
and (as far as possible) updated physical prescriptions is mandatory
in this process. However, even with the best models one should keep in
mind that:
	\begin{itemize}
	\item
Even for a given value of metallicity and age, cluster-to-cluster
variations in the number/distribution of evolved stars lead to
non-neglegible stochastic fluctuations in the colours and hence
to errors in the estimated ages.
	\item
Model colours may present systematic off-sets of some few hundredths
of magnitude, due e.g. to the use of filter pass-bands different from
those used to collect the data.
	\item
Some filter sets are expected to be either (i) better in separating
clusters according to their age and metallicity, or (ii) less
sensitive to the potential errors in the models, and shall hence be
preferred.
	\item
The mass-to-light ratio is determined, primarily, by the choice
of the IMF and of its normalisation. Without a convenient check of
this normalisation, model mass-to-light ratios (and mass estimates)
are meaningless.
	\end{itemize}
We are producing theoretical tools that can be used to face these
problems. This contribution gives a very short anticipation about
them.

\section{Isochrones in several photometric systems}

\begin{figure}
\begin{minipage}{0.48\textwidth}
\psfig{file=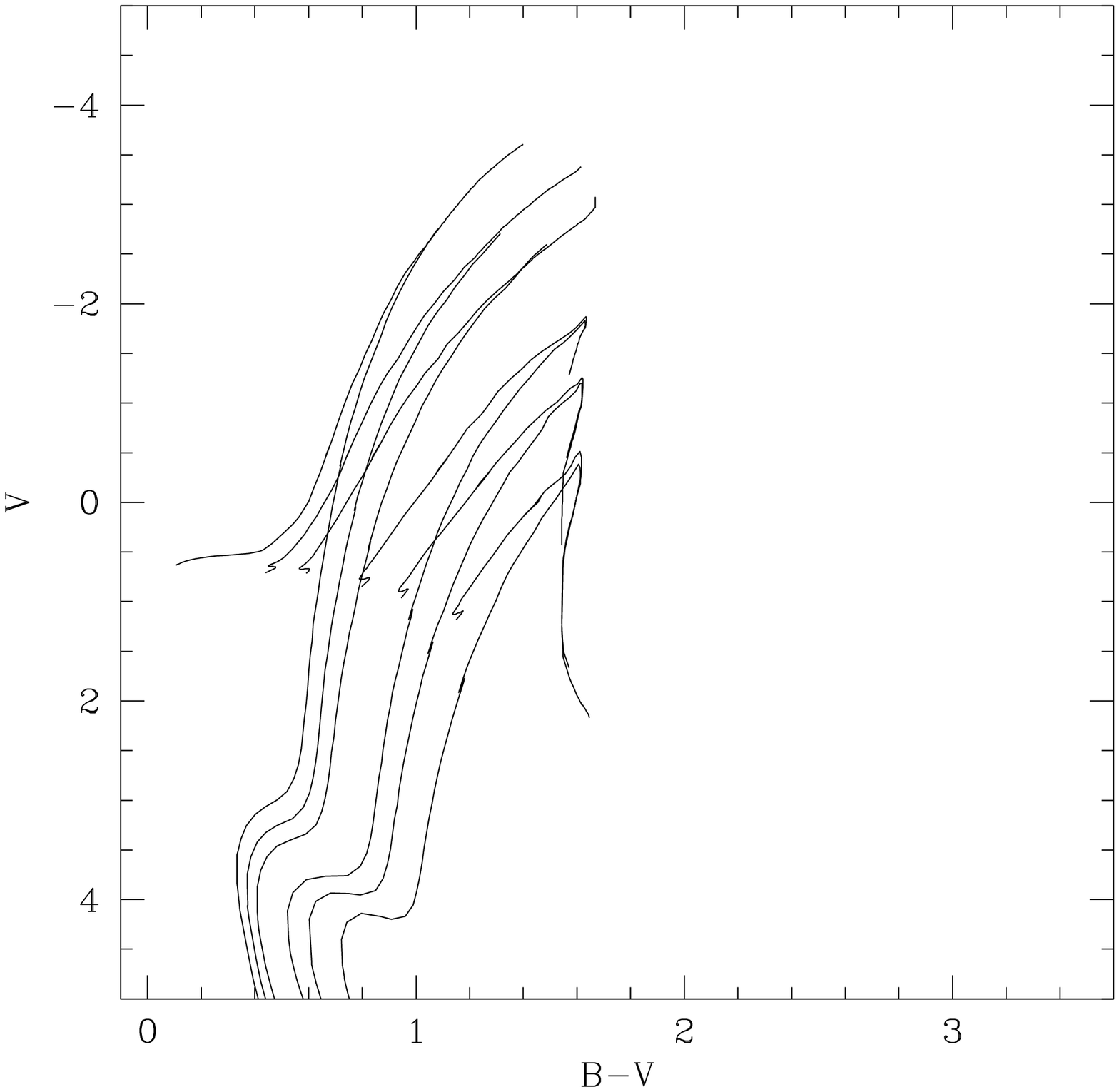,width=\textwidth}
\end{minipage}
\hfill
\begin{minipage}{0.48\textwidth}
\psfig{file=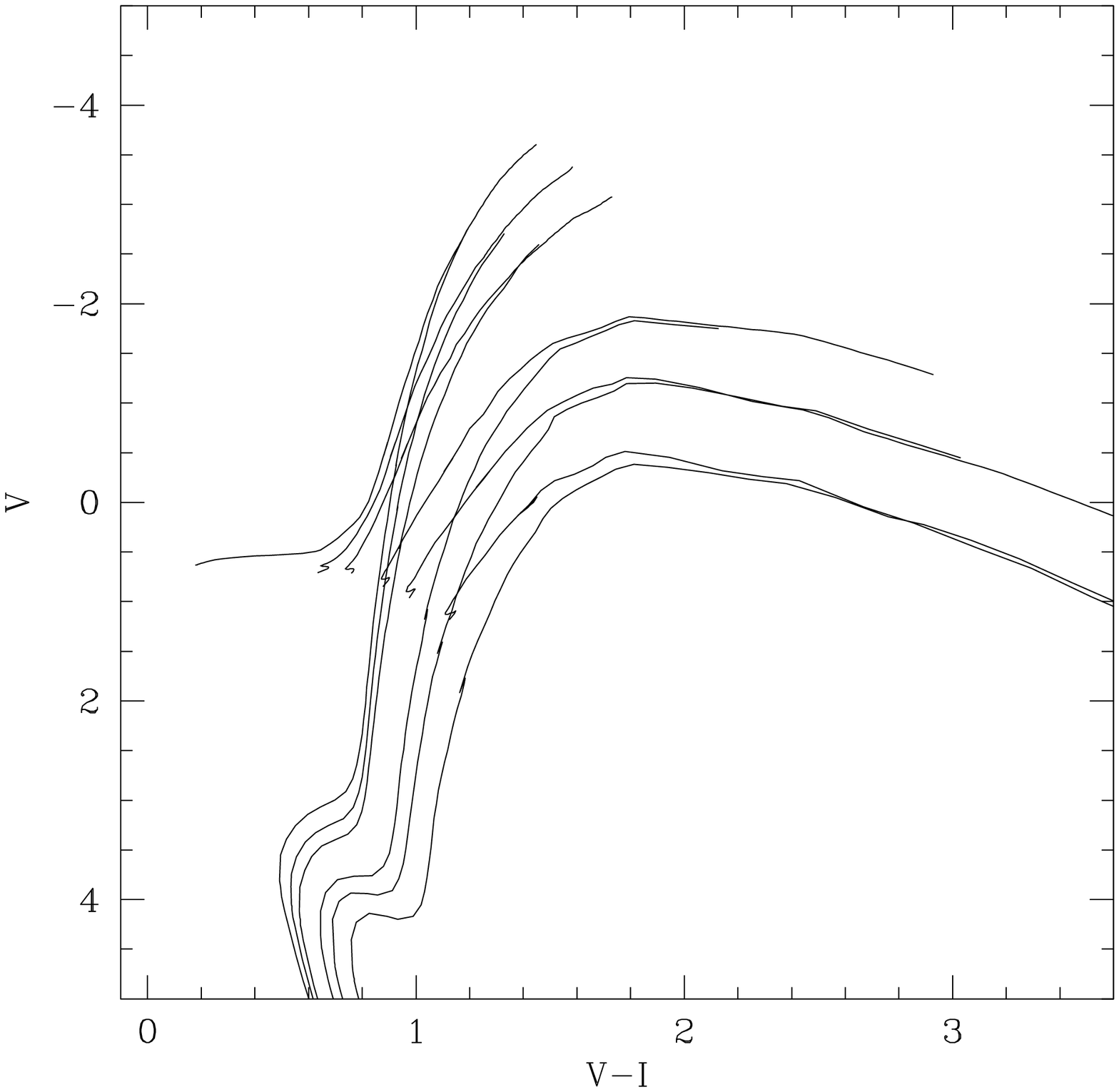,width=\textwidth}
\end{minipage}
\hfill
\begin{minipage}{0.48\textwidth}
\psfig{file=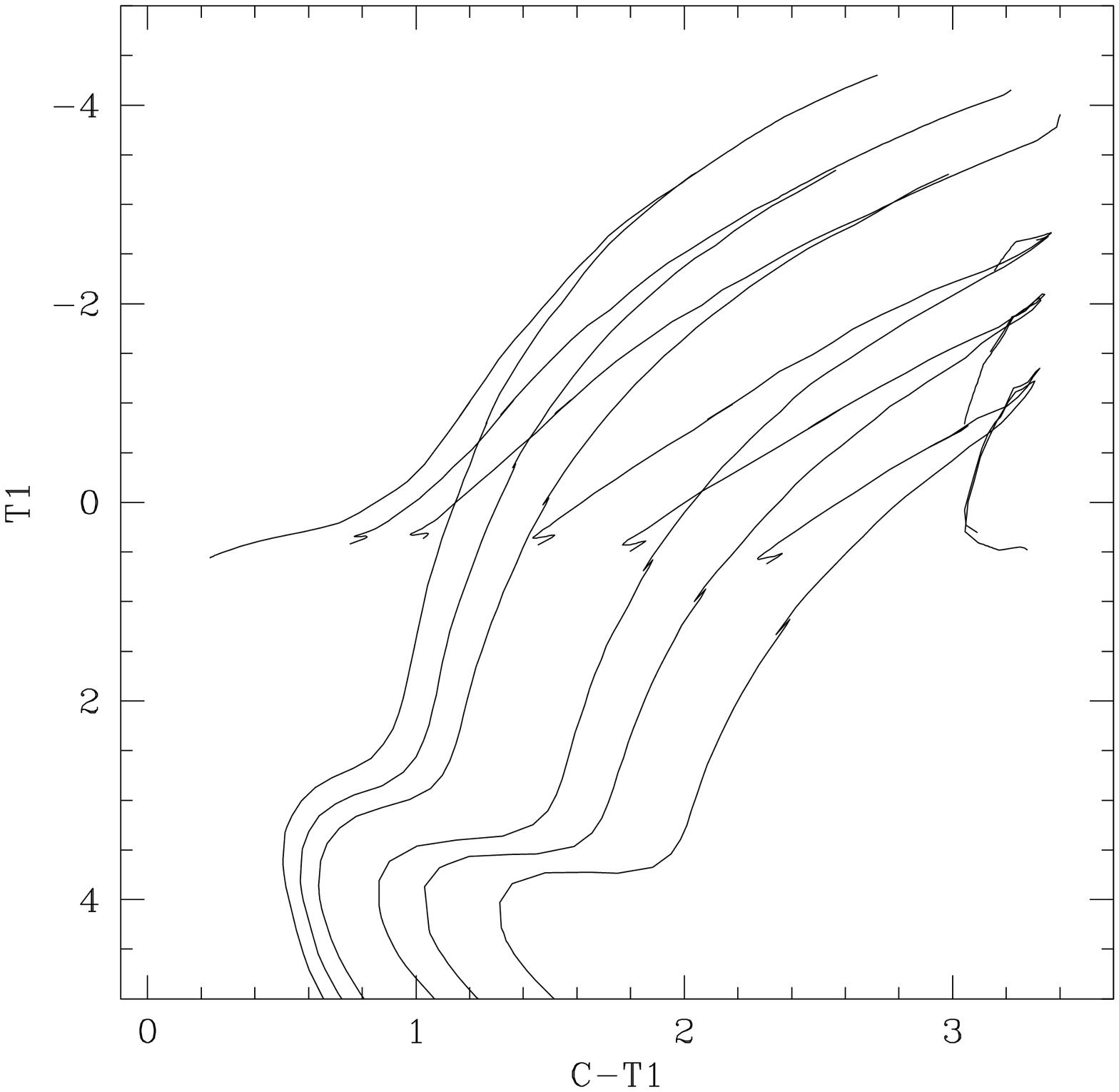,width=\textwidth}
\end{minipage}
\hfill
\begin{minipage}{0.50\textwidth}
\caption{Comparison between Johnson-\-Cou\-sins\- ($BVI$) and
Washington ($CT1$) colour-magnitude diagrams, for a series of 12-Gyr
old isochrones from Girardi et al.\ (2000). In each panel, the
metallicities go from $-2.3$ to $0.0$ (from left to right).}
\label{fig_bvct1}
\end{minipage}
\end{figure}

Our ``basic set'' of stellar evolutionary tracks and isochrones is
presently composed by the Girardi et al.\ (2000) tracks for low- and
intermediate-mass stars, complemented with Bertelli et al.\ (1994, and
references therein) for the massive stars.  For some metallicity
values, we have the possibility of including the more sophisticated
models for the TP-AGB phase from Marigo (2001, and references therein;
Marigo \& Girardi 2001), or the extensive sets of tracks with an
enhanced partition of $\alpha$-elements from Salasnich et al.\
(2000). Some non-overshooting isochrones are available, allowing a
comparison with other classical models from the
literature. Aditionally, we now have new sets with $Z=0.0001$ (yet
unpublished) and $Z=0$ (Marigo et al.\ 2001).

We are now transforming these isochrone sets to several broad-band
photometric systems (Girardi et al.\ 2001). To do so, we put together
an extended library of stellar instrinsic spectra, which is based
essentially on the synthetic atmospheres from Bessell et al.\ (1998)
and Chabrier et al.\ (2000), and on the empirical data for M-giants
from Fluks et al.\ (1994).  Then, we perform synthetic photometry in
several broad-band filter systems, and check for the zero-points. The
resulting tables of bolometric corrections are used to transform our
isochrone sets to observational quantities.

The process has been concluded for the Johnson-Cousins-Glass,
Washington, Thuan-Gunn, HST/WFPC2 and HST/NICMOS systems, and for the
filter sets used by the ESO Imaging Survey (i.e.\ EMMI, SOFI, and WFI
cameras). The same is to be done for the Sloan Digital Sky Survey
system, and others.  In all cases, the only time-consuming task is the
checking of zero-points. Given enough information (filter+detector
transmission curves and zero-point definition), we are able to provide
the same isochrones in {\em any} photometric system upon request.

In Fig.~\ref{fig_bvct1} we provide one example of the utility of this
database: We compare the same set of isochrones of varying
metallicity, in the Johnson-Cousins $B-V$ and $V-I$ colours, and in
the Washington $C-T1$ one. It is evident that $C-T1$ provides the best
separation (at least twice larger than in $B-V$) in metallicity over
the complete magnitude interval, from the main sequence up to the
upper RGB. And, importantly, all these filter pairs have similar
throughputs. This confirms that the integrated $C-T1$ is an excelent
metallicity indicator for old clusters, in accordance to what
indicated by Geisler \& Sarajedini (1999).  We did not find a
comparable performance in other broad-band colours.

\section{Simulations of star clusters}

\begin{figure}
\begin{minipage}{0.48\textwidth}
\psfig{file=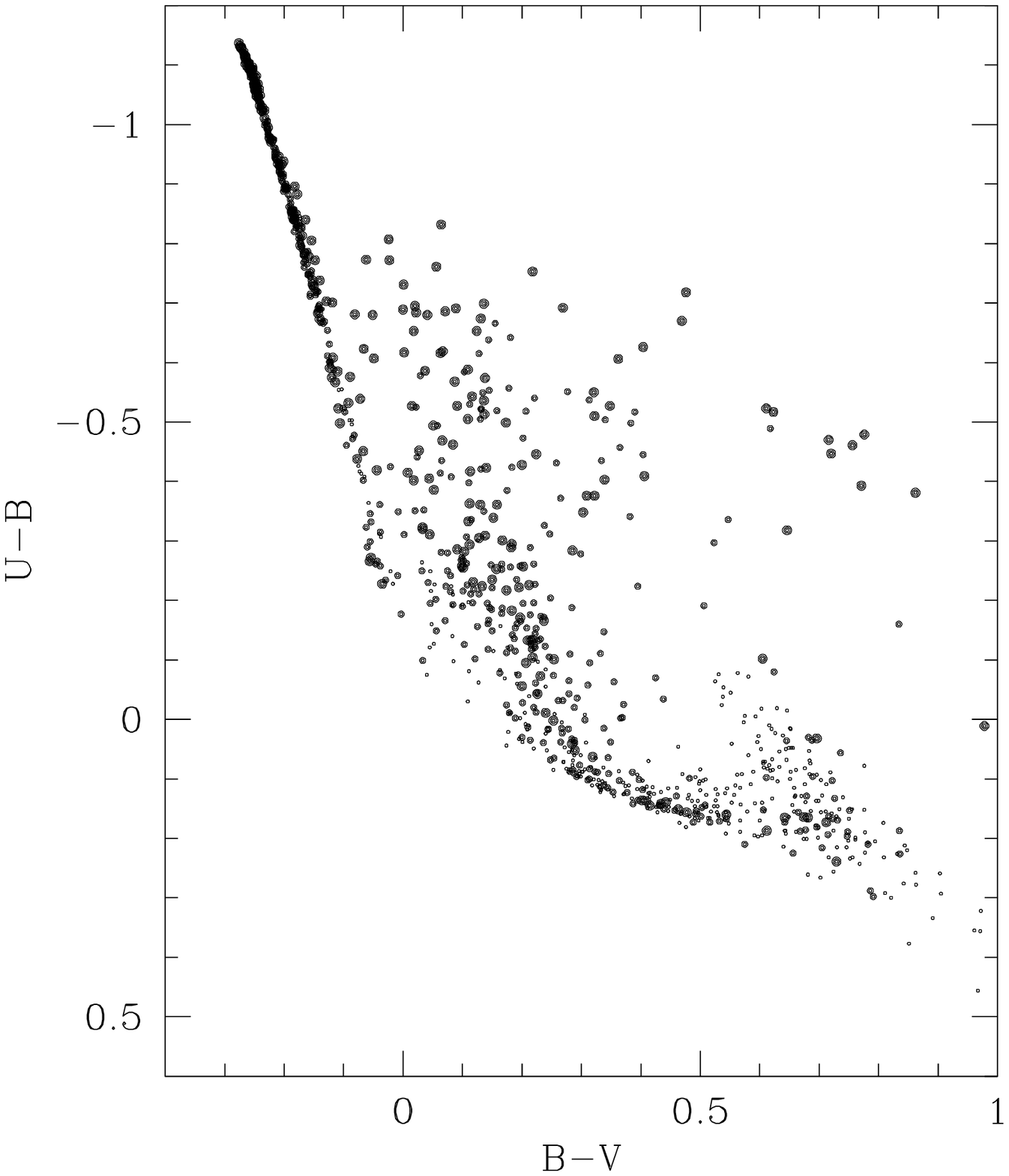,width=\textwidth}
\end{minipage}
\hfill
\begin{minipage}{0.48\textwidth}
\psfig{file=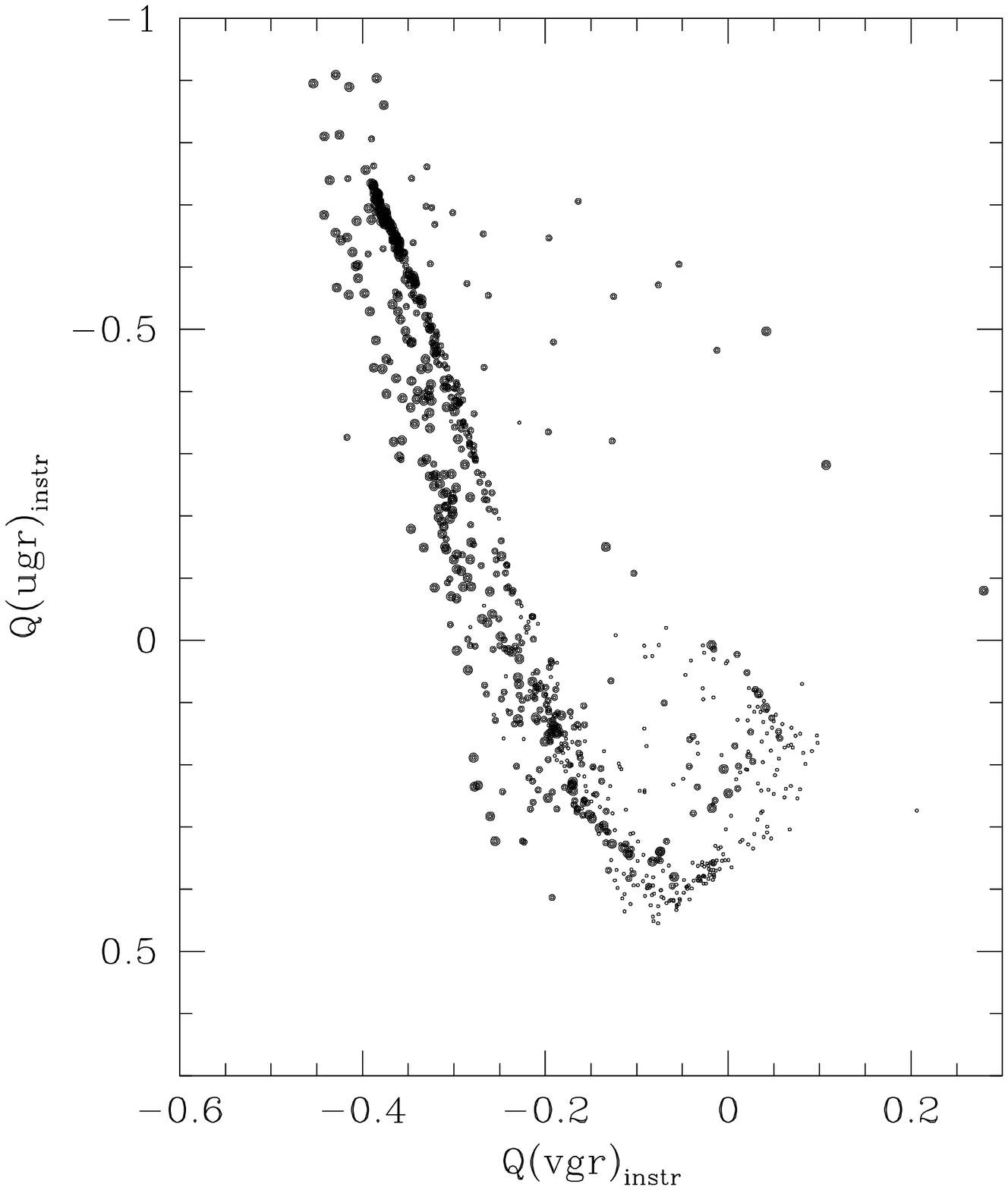,width=\textwidth}
\end{minipage}
\caption{Simulations of a cluster population that obeys a $M^{-2}$
distribution of initial masses, a $t^{-1}$ distribution of ages, a
linear $[{\rm Fe/H}](t)$ age-metallicity relation, with 600 clusters
above the assumed $M_V=5$ limiting magnitude. In the left panel, we
have the $UBV$ plane of Johnson photometry, in the right one the
$Q(ugr)$--$Q(vgr)$ diagram of Thuan-Gunn. See the text for details. }
\label{fig_sim}
\end{figure}

Once we assume an initial mass function (IMF), we can easily compute
either (1) the integrated magnitudes for single-burst stellar
populations, or (2) Monte-Carlo simulations of clusters of given age,
metallicity, and mass (or luminosity).  Since the number ratio of
stars above and below 1~\Msun\ determines, in practice, the
mass-to-light ratios in the models, the IMF should be properly
normalized and calibrated in order to give reasonable values to this
quantity. For a Salpeter IMF, a suitable low-mass cut-off has to be
assumed (see Girardi \& Bica 1993; Girardi et al.\ 1995). If we use the
more recent Kroupa (2001) IMF, $M/L_V$ come out already comparable to
those observed in young LMC clusters, without the need of imposing a
low-mass cut-off. 

Together with the isochrones, we intend to provide the tools for
simulating the integrated light of star clusters, for any assumed IMF,
and as a function of age, metallicity, and total mass (or total
luminosity).  Needless to say, other effects such as reddening and
photometric errors can be simulated as well. Such simulations show the
natural colour (and magnitude) dispersion that derives from stochastic
variations in the number of evolved stars, and are obviously useful
for the interpretation of extragalactic cluster data.

Examples of cluster simulations are presented in the panels of
Fig.~\ref{fig_sim}. They show a synthetic cluster population intended
to mimic the LMC one, but obeying very simple distributions of their
properties: we adopt a $M^{-2}$ distribution of initial masses, a
$t^{-1}$ distribution of ages, a linear $[{\rm Fe/H}](t)$
age-metallicity relation, with 600 clusters above the assumed $M_V=5$
limiting magnitude. The results are plotted both in the $U-B$ vs.\ $B-V$
plane of Johnson photometry (left panel), and in the
$Q(ugr)$--$Q(vgr)$ diagram of Thuan-Gunn (right one). We recall that
these two colour-colour diagrams became classical in the
interpretation of LMC cluster data: the first one defines the Elson \&
Fall (1985) age-dating method (see also Girardi et al.\ 1995), whereas
the second one defines the SWB (Searle et al.\ 1980) classification
scheme.

Although the general distribution of simulated points is similar to
the observed ones for LMC clusters (see the counterparts in Bica et
al.\ 1996, and Searle et al.\ 1980), a detailed comparison reveals
that the assumed distributions are inapropriate.  In particular, the
age distribution $\propto t^{-1}$ causes an excess of clusters in some
regions of the 2-colour planes. In this case, a detailed comparison
with Bica et al.\ (1996) data would allow the derivation of a more
realistic age distribution function (see Girardi et al.\ 1995 for an
alternative approach).

Part of the models here mentioned are already available, in a
preliminar form, at
\verb$http://pleiadi.pd.astro.it/~lgirardi/faq.html$. We are now
working in a user-friendly interface, that would accompany the release
of the complete database.  All details are provided in forthcoming
papers (starting with Girardi et al.\ 2001).

\acknowledgements
I greatly acknowledge the LOC for the travel grant that allowed me to
attend this IAU Symp.  Many thanks are due to G.\ Bertelli, A.\
Bressan, C.\ Chiosi, M.\ Groenewegen, P.\ Marigo, B.\ Salasnich, R.\
Tantalo and A.\ Weiss, my collaborators in the isochrone work, and to
E.\ Bica, D.\ Geisler, E.\ Grebel, J.\ Holtzman for useful
conversations.  This work is funded mainly by the Italian MURST.  The
text was written during a stay at MPA supported by the TMR grant
ERBFMRXCT960086.

\end{document}